\documentclass[useAMS
,usenatbib
]{mnras}

\usepackage[usenames,dvipsnames]{xcolor}

\usepackage{amssymb, amsmath}
\usepackage{epsf}
\usepackage{bm}

\usepackage{natbib}

\usepackage{graphicx}
\usepackage{cancel}

\def\la{\; \raise0.3ex\hbox{$<$\kern-0.75em\raise-1.1ex\hbox{$\sim$}}\;}
\def\ga{\;  \raise0.3ex\hbox{$>$\kern-0.75em\raise-1.1ex\hbox{$\sim$}}\;}

\title[Rotation-induced deep crustal heating of millisecond pulsars]
{Rotation-induced deep crustal heating of millisecond pulsars}

\author[M. E. Gusakov, E. M. Kantor, A. Reisenegger]
{M. E. Gusakov$^{1,2}$, 
E.~M.~Kantor$^{1}$,  
A.~Reisenegger$^{3}$
\vspace{0.5cm}
\\
$^1$Ioffe Physical-Technical Institute of the Russian Academy of
Sciences,
Polytekhnicheskaya 26, 194021 St.-Petersburg, Russia
\\
$^2$Peter the Great St.Petersburg Polytechnic University,
Polytekhnicheskaya 29, 195251 St.-Petersburg, Russia
\\
$^3$Instituto de Astrof{\'\i}sica, Facultad de F{\'\i}sica,
Pontificia Universidad Cat\'olica de Chile,
Av. Vicu\~na Mackenna 4860, 7820436 Macul,\\ Santiago, Chile
}

\begin{document}

\date{Accepted 2015 xxxx. Received 2015 xxxx;
in original form 2015 xxxx}

\pagerange{\pageref{firstpage}--\pageref{lastpage}} \pubyear{2015}

\maketitle

\label{firstpage}

%
\begin{abstract}
The spin-down of a neutron star, e.g. due to magneto-dipole losses,
results in compression of the stellar matter and induces nuclear reactions at phase transitions
between different nuclear species in the
crust. We show that this mechanism is effective in heating recycled pulsars, in which the previous accretion process has already been compressing the crust, so it is not in nuclear equilibrium.
We calculate the corresponding emissivity and confront it with available observations, showing that it
might account for the likely thermal ultraviolet emission of PSR~J0437$-$4715.
\end{abstract}
%


\maketitle

\section{Introduction}

The detection of likely thermal ultraviolet emission from the
millisecond pulsar PSR~J0437$-$4715 (\citealt*{kargaltsev04}; \citealt{durant12})
indicates that some reheating mechanisms 
operate in old
pulsars.
\cite*{reisenegger10} reviewed various possibilities, concluding that the most effective ones are
vortex creep (\citealt*{alpar84}) and rotochemical heating (\citealt*{reisenegger95}), both
very sensitive to nucleon superfluidity parameters.
Although the theoretical and observational uncertainties are difficult to quantify (Gonz\'alez-Caniulef \& Reisenegger, in preparation), calculations by
\cite*{pr10} and \cite*{reisenegger15} suggest that
rotochemical heating can explain the observed temperature of
PSR~J0437$-$4715 only if
either the proton or neutron energy gap is sufficiently large
throughout the whole stellar core, which appears to be unlikely,
at least for massive neutron stars (NSs).
Here we discuss a variant of
rotochemical heating that operates in the {\it crusts} of pulsars that were previously recycled through accretion. In these, the accreted matter has been compressed slowly, at moderate temperatures, and therefore its nuclear transformations have not been able to reach its ground state of fully catalyzed matter \citep*{hz90}. Thus, a small additional compression, now due to the decreasing centrifugal force, can immediately induce further reactions at the interfaces between layers of different dominant nuclei%
\footnote{The very similar process called `deep crustal heating' occurs during the accretion process,
in which the matter falling onto the NS surface compresses the underlying layers
of the crust; see, e.g., \citealt*{hz90a,bbr98}.}.
This process provides an additional heating source that has been neglected
in the literature and is considered here.
We emphasize that the proposed heating mechanism is not relevant for fully catalyzed (equilibrium) NS crusts, which
need to be compressed
substantially in order to initiate these transformations.
This explains why \cite*{is97}
found a much lower heating rate than
that obtained in the present paper
(see also footnote 2).

In Section \ref{mechanism} we discuss how the compression
produced as the star spins down initiates nuclear reactions
at isobaric and isopycnic phase transition surfaces in the crust, leading to energy release and heating.
In Section \ref{evaluating}, we present three ways of calculating the heating rate, at different levels of approximation: a simple, Newtonian estimate, a slow-rotation approximation within General Relativity, and a general-relativistic, numerical calculation valid also at fast rotation rates.
Section \ref{results} compares the three approaches,
presents the results and analyzes the role of the proposed new heating mechanism
in the interpretation of observations of the pulsar PSR~J0437$-$4715.
We conclude in Section \ref{concl}.

\section{Crust compression and heating in a spinning-down NS}
\label{mechanism}

As a NS spins down, the compression and shear components of its deformation tensor are of the same order of magnitude, but the shear modulus of the crust is much smaller than the bulk modulus (e.g., \citealt*{hpy07}), allowing us to neglect shear stresses and treat the whole star as a fluid. In the rotating star, the condition of hydrostatic equilibrium requires isobaric (constant pressure $P$) and isopycnic (constant density $\rho$) surfaces to coincide (see, e.g., \citealt*{fr05}). Since the relation between pressure and density depends on chemical composition, the latter must also be uniform on each of these surfaces, but can change from surface to surface. Thus, as the star spins down, any such surface will deform (becoming more spherical) and the associated pressure and density will increase, but the same particles must remain associated with it (otherwise the chemical composition would become non-uniform on a given isobaric surface as an NS decelerates)%
\footnote{
This contradicts a result of \cite{is97}, who found
that the stellar matter compresses at the equator and expands at the poles.
This suggests that there is some mistake in their
approach, most probably in their calculations
of the Lagrangian displacements
(see their equation 52).}.
%

In the crust, there are various phase transitions (which we label by an index $i$), defined by critical pressures $P_i$. When
the pressure of a given isobaric surface reaches $P_i$,
nuclei at this surface undergo
exothermic reactions
(electron captures, neutron emissions, or pycnonuclear reactions), releasing an energy $q_i$ per nucleon. If the total number of nucleons at $P<P_i$ is $\Delta N_i$, and the angular velocity $\Omega$ of the star decreases, the number of nucleons crossing the phase transition $i$ per unit time is $-(d\Delta N_i/d\Omega)\dot\Omega$, where we use the convention that total derivatives correspond to stellar models with the same total baryon number $A$ (but generally different rotation rates) and dots denote time derivatives. The total energy released per unit time at all these phase transitions, redshifted to a distant observer, is
\begin{equation}
\dot{E}^\infty
=-\sum_i \frac{d\Delta N_i}{d\Omega} \, \,
\frac{\dot\Omega\, q_i}{u^t_{(i)}}.
\label{Einfty}
\end{equation}
The gravitational redshift correction is the inverse of the time component of the four-velocity $u^\mu$ of a fluid element \citep*{mrl93},
\begin{equation}
u^t=\left(-g_{tt}-2 \Omega g_{t\phi} -\Omega^2 g_{\phi \phi} \right)^{-1/2}
\label{ut}
\end{equation}
where $g_{tt}$, $g_{t \phi}$, and $g_{\rm \phi \phi}$ are components of the metric of the rotating star.
It can be shown that $u^t$ is also uniform on isobaric surfaces \citep*{fr05}, so its value $u^t_{(i)}$ at a given transition $i$ is well-defined.

\section{Evaluating the heating rate}
\label{evaluating}
\subsection{A simple Newtonian estimate}

Applying the Newtonian condition of hydrostatic equilibrium to the (assumed) thin layer of matter above the transition $i$ in a non-rotating star, we can write
\begin{equation}
\Delta N_i(\Omega=0)\approx \frac{4\pi P_iR^2}{m_{\rm B}g}\approx \frac{4\pi P_iR^4}{G M m_{\rm B}},
\label{N0}
\end{equation}
where $R$, $M$, and $g$ are the stellar radius, mass, and surface gravity, $G$ is the gravitational constant, and $m_{\rm B}$ is the mass of a free nucleon (taken to be the same for neutrons and protons).
When the star rotates slowly ($\Omega\ll \Omega_{\rm K}$, where $\Omega_{\rm K}\sim(GM/R^3)^{1/2}$ is the Kepler frequency or break-up spin rate), there will be a correction:
\begin{equation}
\Delta N_i(\Omega)
\approx\Delta N_i(0)\left[1+a\left(\frac{\Omega}{\Omega_{\rm K}}\right)^2\right],
\label{exp}
\end{equation}
with $a$ being a positive coefficient of the order of unity.
Thus,
\begin{equation}
-\frac{d \Delta N_i}{d\Omega}\dot\Omega\approx
-2 a \, \frac{\Omega \dot{\Omega}}{\Omega_{\rm K}^2} \, \Delta N_i(0)
\approx -a \, \frac{8\pi R^7}{G^2 M^2 m_{\rm B}}
\Omega \dot{\Omega}
\, P_i,
\label{dN2}
\end{equation}
from which we get the approximate expression
\begin{equation}
\dot{E}
\approx -a \, \frac{8\pi R^7}{G^2 M^2 m_{\rm B}} 
\Omega \dot{\Omega} 
\sum_i \, P_i
\, q_i,
\label{Efit}
\end{equation}
where we also suppressed the redshift factor 
because we ignored the effects of general relativity. 
Also, for simplicity, we assumed the same value of $a$ for all phase transitions.
Scaling with fiducial values $R_6=R/(10^6\,\rm cm)$, $M_{1.4}=M/(1.4M_\odot)$,
$\Omega_{3}=\Omega/(10^3\,\rm s^{-1})$,
and $\dot{\Omega}_{-14}=\dot \Omega/(10^{-14}\,\rm s^{-2})$,
equation (\ref{Efit}) becomes
\begin{eqnarray}
\dot{E}
&\approx& -7\times 10^{28} \, a\,
\frac{R_6^7}{M_{1.4}^2}
\, \Omega_{3} \, \dot{\Omega}_{-14}\,
\nonumber\\
 &\times& \sum_i
\frac{P_i}
{10^{31}\,\rm erg\,cm^{-3}}
\,
\frac{q_{i}}{1\,\rm MeV}
\,\, {\rm erg}\, {\rm s}^{-1}.
\label{Efit1}
\end{eqnarray}
Notice the strong dependence of the heating rate
(and thus the bolometric luminosity)
on the stellar radius ($\propto R^7$).
This is reduced, however, when considering the effective temperature,
$T_{\rm eff}\propto(\dot E/R^2)^{1/4}\propto R^{5/4}$,
or the Rayleigh-Jeans flux (relevant in the ultraviolet region),
$F_{\mathrm RJ}\propto R^2T_{\rm eff}\propto R^{13/4}$.

It follows from equation (\ref{Efit1}) that the heating rate generally depends
only on 4 ``microphysics-related'' parameters, $M$, $R$, $P_i$, and $q_i$
(and this is verified by the results of numerical analysis in Section 4).
The first two 
($M$ and $R$)
are mainly determined by the core equation of state (EOS),
while the second pair ($P_i$ and $q_i$) depends on the (rather uncertain)
details of the accreted crust EOS (see Section 4 for a more detailed discussion).

\begin{figure*}
    \begin{center}
        \leavevmode
        \epsfxsize=6in \epsfbox{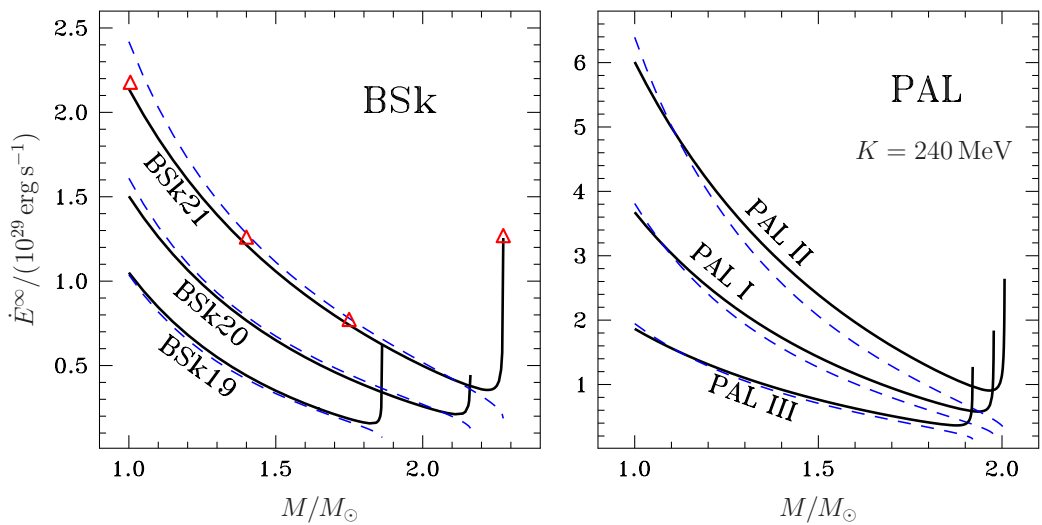}
    \end{center}
    \caption{
    Heating rate in the crust normalized to $10^{29}\,\rm erg\,s^{-1}$
		as a function of NS mass for three BSk EOSs (left panel) and three PAL EOSs (right panel).
		Thick solid lines are calculated in the approximation of slow rotation,
		using formula (\ref{Einfty}) and (\ref{Ndot1}).
		Thin dashed lines represent the approximation in equation (\ref{Efit1}) with the
        choice $a=1/2$.
		Triangles show the heating rate, calculated for the BSk21 EOS
        with the full numerical procedure,
        described in Section \ref{rapid}.
		All curves are plotted for $\nu=100\,\rm Hz$
		and $\dot\Omega=-10^{-14}\,\rm s^{-2}$.}
    \label{Fig:Emiss}
\end{figure*}

\subsection{General-relativistic slow-rotation approximation}
\label{hartle}

Still in the limit of slow rotation, $\Omega/\Omega_{\rm K} \ll 1$, we now aim at a more precise evaluation including the effects of General Relativity.
When an NS decelerates ($\Omega$ decreases),
its central density $\rho_{\rm c}$,
given by
$\rho_{\rm c}(\Omega)=\rho_{\rm c0}+\delta \rho_{\rm c}(\Omega)$,
increases ($\rho_{\rm c0}$ is the central density at $\Omega=0$;
$\delta \rho_{\rm c}$ is the small $\Omega$-dependent correction).
The correction $\delta \rho_{\rm c}$
can be found from the condition that the total number of baryons in the star,
$A(\rho_{\rm c},\Omega)$,
remains unchanged as the star spins down,
\begin{equation}
A(\rho_{\rm c0}+\delta \rho_{\rm c}, \, \Omega)=A(\rho_{\rm c0}, \, 0).
\label{AA}
\end{equation}
Expanding the left-hand side of (\ref{AA})
in a Taylor series,
one arrives at the following quadratic equation for $\delta \rho_{\rm c}(\Omega)$,
\begin{equation}
\beta \, \delta \rho_{\rm c} + \gamma \, \delta \rho_{\rm c}^2 + \alpha \, \Omega^2 \approx 0,
\label{AA2}
\end{equation}
where $\beta=\partial A/\partial \rho_{\rm c}$,
$\gamma=(1/2)\partial^2 A/\partial \rho_{\rm c}^2$,
$\alpha =(1/2)\partial^2 A/\partial \Omega^2$ and
all derivatives are taken at $\rho=\rho_{\rm c0}$ and $\Omega=0$.
To derive (\ref{AA2}) we used the fact that $\partial A/\partial \Omega \propto \Omega$
and hence vanishes at $\Omega = 0$.
Knowledge of $\delta \rho_{\rm c}(\Omega)$
(for more detailed analysis of equation \ref{AA2} see below)
allows us to write
\begin{equation}
\frac{d \Delta N_i
}{d\Omega}
=
\frac{\partial \Delta N_i
}{\partial \rho_{\rm c}}\frac{d\rho_{\rm c}}{d\Omega}+\frac{\partial \Delta N_i
}{\partial \Omega}
=\frac{\partial \Delta N_i
}{\partial \rho_{\rm c}}\frac{d\delta\rho_{\rm c}}{d\Omega}+\frac{\partial \Delta N_i
}{\partial \Omega}.
\label{Ndot1}
\end{equation}
We evaluate the latter expression using the scheme of \cite{hartle67} and \cite{ht68}, following the detailed description of \cite{fr05} and making use of
our equation (\ref{AA2}) and
their equations (18)--(26), which we checked by an independent calculation.

In the limit $\Omega\to 0$, it is expected that $\partial\Delta N_i/\partial\rho_c\to\rm{constant}$ and $\partial\Delta N_i/\partial\Omega\propto\Omega$, which we
confirmed by numerical evaluations.
On the other hand,
$d\delta \rho_{\rm c}/d\Omega$
has a less simple behaviour.
As follows from equation (\ref{AA2}),
if $\beta$ is not small then the term $\gamma \, \delta \rho_{\rm c}^2$
can be neglected and
$\delta \rho_{\rm c} \approx -\alpha \Omega^2/\beta \propto \Omega^2$.
As a result, $\dot{E}^\infty\propto\Omega\dot\Omega$ at $\Omega \to 0$, as in equations (\ref{Efit}) and (\ref{Efit1}).
However,
near the maximum (non-rotating) NS mass $M_{\rm max}$, which occurs at the same central density, $\rho_{c,\rm max}$, as the maximum baryon number, $A_{\rm max}$ (e.g., \citealt*{hpy07}),
the situation is different.
The coefficient $\beta$ there is small
(and vanishes at $\rho_{\rm c0}=\rho_{c,{\rm max}}$)
so that the terms $\propto \delta \rho_{\rm c}$ and $\delta \rho_{\rm c}^2$
in equation (\ref{AA2})
can be of comparable magnitude.
In the most extreme case ($\beta=0$)
one obtains $\delta \rho_{\rm c} = -(-\alpha \Omega^2/\gamma)^{1/2}  \propto \Omega$
so that $\dot{E}^\infty\propto\dot\Omega$
at $\Omega \to 0$ rather than $\propto \Omega\dot\Omega$.

\subsection{General-relativistic, rapidly rotating NSs}
\label{rapid}
This problem was also studied numerically, without reference to a slow-rotation approximation,
using the RNS code developed by S. Morsink and N. Stergioulas
({\bf http://www.gravity.phys.uwm.edu/rns/}).
This code constructs relativistic models of rapidly rotating NSs
for a given tabulated equation of state.

We calculate a set of NS models for the same total baryon number $A$ and different
spin rates $\Omega$.
Then we look for the number of baryons located above the phase transition $i$
making use of equation (20) of \cite{fr05}, and approximate its dependence on $\Omega$
by an analytic function.
Differentiating this fit,
we calculate $d \Delta N_i
/d\Omega$, to be used in
equation (\ref{Einfty}).

We follow the same strategy as in the slow-rotation limit (see Section \ref{hartle}), 
using equation (\ref{Einfty}) with 
$u^t_{(i)}=(-g_{tt})^{-1/2}$ evaluated for
a non-rotating star. This approximation appears to be rather accurate even for the most rapidly rotating NSs,
not deviating
from the accurate calculation
by more than a couple per cent, as confirmed by
\citet{bopm15}.

\section{Results}
\label{results}

To illustrate our results, we employ
three EOSs of the BSk (Brussels-Skyrme) family
(BSk19, BSk20, and BSk21;
see \citealt*{pfcpg13} for details),
and three stiff
(compression modulus $K=240 \,\rm MeV$) EOSs from \cite*{pal88};
hereafter PAL EOSs (PAL I, PAL II, and PAL III) to describe matter in the NS core%
%
\footnote{Softer EOSs are inconsistent with the precisely measured NS masses around $2 M_\odot$ \citep{demorest10,antoniadis13}.}.
%
The unified BSk EOSs describe NS matter
in the whole range of densities including the core and crust.
However, \cite{pfcpg13} do their calculations only for catalyzed matter of NS crust,
while here we are interested in NSs
with an accreted crust that has not relaxed to the nuclear equilibrium state.
Thus, although the
actual functional
dependence of $P$ on $\rho$
in the crust
has little effect on our results
(because the heating rate depends on $M$ and $R$, see equation \ref{Efit1},
which are mainly determined
by the function $P(\rho)$ in the core and not in the crust),
for
consistency we employ an EOS of accreted crust from \cite{hz90}, which we match
with
the BSk or PAL EOSs in the core.
On the other hand, 
the 
pressure at the phase transitions ($P_i$)
and the heat release per baryon $q_i$ 
in nuclear transformations at these transitions
depend essentially on  details of EOS in the accreted crust;
they were taken
from Table A.3 of \cite{hz08},
thus assuming that X-ray ashes consist of pure $^{56}{\rm Fe}$.
Note that, as is evident from the above discussion,
the EOSs employed by us in the core and in the accreted crust
have not been derived from the same microphysics input.
Such calculations, which would allow one to determine
{\it self-consistently} the parameters
$M$, $R$, $P_i$, and $q_i$,
are still unavailable in the literature
(but one can make a rough guess about the possible uncertainty involved into the problem;
see 
footnote 4 and a discussion below).

Figure \ref{Fig:Emiss} allows to compare the results for all three approaches described in the previous section.
As it should be for a rather slowly rotating star
($\nu=100\,\rm Hz$), the numerical results (triangles),
calculated as discussed in Section \ref{rapid},
almost coincide with the slow-rotation approximation (thick curves).
Even the rough estimate from equation (\ref{Efit1}) (thin dashed curves) follows them quite well, only failing to reproduce the sharp upturn close to the maximum mass, due to the anomalous behaviour of the function $d \Delta N_i/d\Omega$ near the maximum mass discussed in Section 3.2, yielding just $\dot E^\infty\propto\dot\Omega$, in this limit, as opposed to $\dot E^\infty\propto\Omega\dot\Omega$ at the lower masses ($M \lesssim 0.95 \, M_{\rm max}$) to which the estimate from equation (\ref{Efit1}) applies.
Keeping these proportionalities in mind,
it is easy to rescale this figure to any values of $\nu \ll \Omega_{\rm K}/(2 \pi)$ and $\dot\Omega$.

As suggested by the estimate of equation (\ref{Efit1})
and confirmed by the calculation results in figure \ref{Fig:Emiss},
the heating rate $\dot E^\infty$ is indeed rather sensitive to
the stellar mass and
the EOS in the NS core (which together determine the stellar radius).
It is also sensitive to the positions of phase transitions
in the crust and energy release due to nuclear transformations
at the phase transitions ($P_i$ and $q_i$, see Eq. \ref{Efit1}).
These quantities are rather model-dependent and, in particular,
depending on the nuclear symmetry energy,
the heating rate $\dot{E}^{\infty}$ can vary
by about a factor of two (\citealt*{steiner12})%
%
%
\footnote{
Actually, what \cite{steiner12} estimated is the uncertainty of the deep crustal heating rate
in accreting NSs, which is $\propto \sum_i q_i$.
In our case, $\dot{E}^{\infty}$ is proportional to $\sum_i P_i q_i$ (see equation \ref{Efit1}),
so that, strictly speaking, the uncertainty can be somewhat different.
However, \cite{steiner12} does not provide any tables with his data for $P_i$ and $q_i$
(as \citealt{hz08} did), so we cannot implement his results directly to check the real uncertainty.}.
%
Taking into account all these uncertainties, $\dot E^\infty$ can be estimated
as $\dot E^\infty\approx(0.5-20)\times 10^{-5}\dot{E}_{\rm rot}$
(where $\dot{E}_{\rm rot}=I\Omega\dot{\Omega}$ is the rotation energy loss; $I$ is the stellar moment of inertia).

\begin{figure}
    \begin{center}
        \leavevmode
        \epsfxsize=3.3in \epsfbox{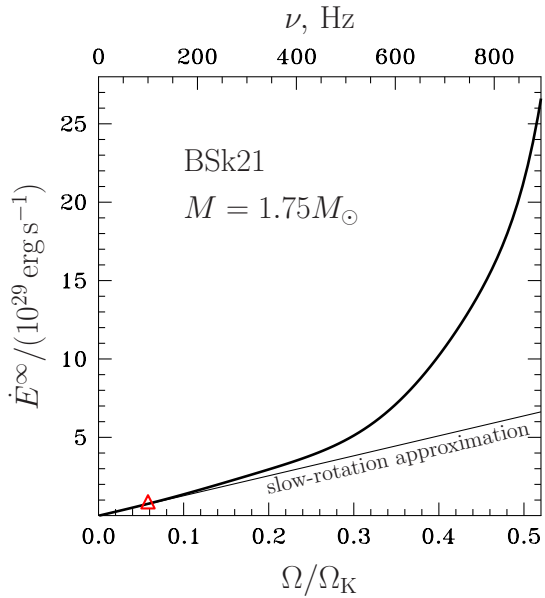}
    \end{center}
    \caption{
     $\dot{E}^\infty$ normalized to $10^{29}$~erg~s$^{-1}$
		as a function of $\Omega$ in units of Kepler frequency $\Omega_{\rm K}$
		for BSk21 EOS and an NS with $M=1.75 M_\odot$
		(according to the RNS code $\Omega_{\rm K}=10807.9\,\rm s^{-1}$).
		The thin curve is calculated in the slow-rotation approximation,
		using formulas (\ref{Einfty}) and (\ref{Ndot1}), while
		the thick curve is calculated with the RNS code (see Section \ref{rapid}).
		The triangle shows $\dot{E}^\infty$ for $\nu=100\,\rm Hz$,
        the frequency used in figure \ref{Fig:Emiss}.
		To plot the figure we take $\dot\Omega=-10^{-14}\,\rm s^{-2}$.
 }
    \label{Fig:EmissOmega}
\end{figure}

Figure \ref{Fig:EmissOmega} illustrates
the accuracy of the slow-rotation approximation as a function of rotation rate.
As expected, for rapidly rotating stars ($\nu\ga 500$~Hz)
the results of these
two approaches differ substantially,
with the exact solution being several-fold larger than the slow-rotation limit predicts.

Surface temperatures of millisecond pulsars are generally not known.
However, \cite{kargaltsev04} and \cite{durant12} have measured likely thermal emission from PSR~J0437$-$4715
in the far ultraviolet, implying that some reheating mechanisms must operate in this pulsar.
The latter authors derive a redshifted effective temperature in the range
$1.25\times 10^5\,\rm K<T_{\rm eff}^\infty<3.5\times 10^5\,\rm K$, under the assumption of a blackbody emitter with redshifted circumferential radius in the range $7.8\,\mathrm{km}<R_\infty<15\,\mathrm{km}$.

\begin{figure}
    \begin{center}
        \leavevmode
        \epsfxsize=3.3in \epsfbox{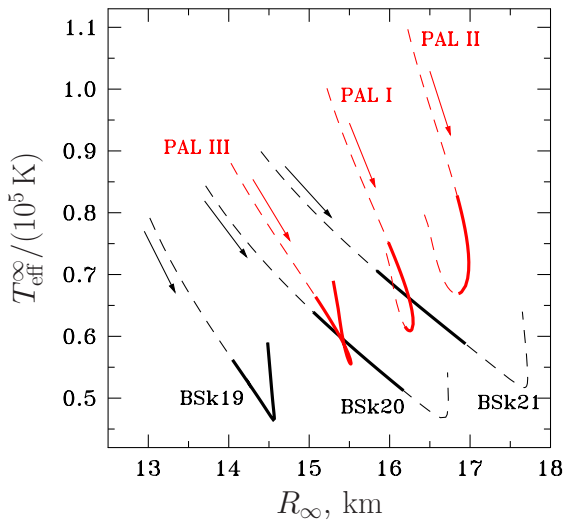}
    \end{center}
    \caption{
    Effective $T_{\rm eff}^\infty$ (in units of $10^5\,\rm K$)
		as a function of redshifted circumferential radius (or ``radiation radius'') $R_\infty$ for six EOSs (see figure \ref{Fig:Emiss}) calculated in the slow-rotation approximation with $\Omega$ and $\dot\Omega$ taken for PSR~J0437$-$4715
		($\Omega=1091\,\rm s^{-1}$, $\dot\Omega=2.60\times 10^{-15}\,\rm s^{-2}$).
		Thin dashes show the mass range $1M_\odot<M<M_{\rm max}$ for each EOS, while thick lines correspond to $1.56M_\odot<M<1.96M_\odot$ (the allowed mass range for PSR~J0437$-$4715 at 68\% confidence as measured by \citealt{verbiest08}). Arrows indicate the direction of increasing mass.
 }
    \label{Fig:RT}
\end{figure}

Figure \ref{Fig:RT} shows the predictions from the reheating mechanism proposed in this paper.
As one can see, rotation-induced heating in the crust is a powerful mechanism, being close to explaining by itself (apart from other heating processes, such as vortex friction or rotochemical heating) the observed temperature of PSR~J0437$-$4715.
Note that the upper limit for $R_\infty=15\,\mathrm{km}$ 
assumed by \cite{durant12} when deriving the temperature from the ultraviolet flux
is smaller than the values obtained for most of the currently considered EOSs, including those used in this paper. Since the ultraviolet flux (in the Rayleigh-Jeans limit of the thermal spectrum) is roughly $\propto R_\infty^2 T_{\rm eff}^\infty$, one could extrapolate the lower limit of their temperature range as $T_{\rm eff}^\infty\approx 1.25\times 10^5(15\,\mathrm{km}/R_\infty)^2\,\mathrm{K}$, yielding $T_{\rm eff}^\infty \approx 0.97\times 10^5\,\mathrm{K}$ at $R_\infty=17\,\mathrm{km}$. Given various uncertainties in both the measurements and the theoretical models, as well as the crude assumption of a blackbody atmosphere, the predictions from our model might be considered to be in marginal agreement with the observation. The agreement might be improved by including some other reheating mechanisms as well.

\section{Conclusions}
\label{concl}

We have shown that the compression of the accreted crust in the course of NS spin-down
results in a significant energy release at the phase transitions in the crust.
The corresponding heating rate, $\dot{E}^\infty$, can be estimated as
$(0.5-20)\times 10^{-5}\dot{E}_{\rm rot}$.
In contrast to the other most effective reheating mechanisms proposed before
--- rotochemical heating (\citealt{reisenegger95}) and vortex creep (\citealt{alpar84}) --- the reported reheating mechanism is not affected by the rather uncertain parameters of baryon superfluidity in the NS.
As we have demonstrated, $\dot{E}^\infty$ mainly depends on the stellar mass and radius, as well as on the positions of the phase transitions in the crust and the energy release at these transitions; see equation (\ref{Efit1}). Noteworthy, it is insensitive to other details of the crust and core EOSs.

The derived
$\dot{E}^\infty$
is one to two orders of magnitude larger
than the values obtained by \cite{is97}
and is comparable to the rotochemical heating rate (\citealt{fr05,reisenegger15}).
For some EOSs, the predicted surface temperature is marginally consistent with the observation of PSR~J0437$-$4715 \citep{durant12}.
However, some other reheating mechanisms could also be important for this pulsar.

\section*{Acknowledgments}
We are very grateful to Sharon Morsink for her
kind help with the RNS code,
to Denis Gonz\'alez-Caniulef
for some useful clarifications concerning PSR J0437--4715,
to
Y.A.~Shibanov, D.G.~Yakovlev, and Dima Zyuzin
for discussions,
to
Oleg Kargaltsev and Martin Durant for correspondence,
and to Andrey Chugunov for helpful comments.
E.M. Kantor was partially supported by
RFBR (grants 14-02-00868-a and 14-02-31616-mol-a)
and by RF president programme
(grants MK-506.2014.2 and NSh-294.2014.2).
A.~Reisenegger was supported by FONDECYT (Chile) Regular Projects 1110213 and 1150411;
a visit of E.M. Kantor and M.E. Gusakov to Chile was
partially supported by FONDECYT (Chile)
Regular Project 1110213.


\label{lastpage}

\end{document}